\documentclass[structabstract]{aa}
\usepackage{amsfonts}

\usepackage{graphicx}
\usepackage{txfonts}

\begin{document}
   \title{Determining gravitational wave radiation from close galaxy pairs using a binary population synthesis approach}

   \author{Jinzhong Liu
          \inst{1},
          Yu Zhang \inst{2,3}, Hailong Zhang \inst{1}, Yutao Sun
          \inst{2}, and Na Wang \inst{1}
          }


   \institute{National Astronomical Observatories/Xinjiang Observatory, the Chinese Academy of Sciences, 150, Science 1-Street Urumqi, Xinjiang 830011
China\\
              \email{liujinzh@xao.ac.cn}
              \and
              National Astronomical Observatories/Yunnan Observatory, Chinese Academy of Sciences, Kunming, 650011,
              P.R. China\
              \and
              Graduate University of the Chinese Academy of Sciences, Beijing 100049, China}

   \date{Received ; accepted}


  \abstract
{ The early phase of the coalescence of supermassive black hole
(SMBH) binaries from their host galaxies provides a guaranteed
source of low-frequency (nHz-$\mu$Hz) gravitational wave (GW)
radiation by pulsar timing observations. These types of GW sources
would survive the coalescing and be potentially identifiable.}
{We aim to provide an outline of a new method for detecting GW
radiation from individual SMBH systems based on the Sloan Digital
Sky Survey (SDSS) observational results, which can be verified by
future observations.}
{Combining the sensitivity of the international Pulsar Timing Array
(PTA) and the Square Kilometer Array (SKA) detectors, we used a
binary population synthesis (BPS) approach to determine GW radiation
from close galaxy pairs under the assumption that SMBHs formed at
the core of merged galaxies. We also performed second post-Newtonian
approximation methods to estimate the variation of the strain
amplitude with time. }
{We find that the value of the strain
    amplitude \emph{h} varies from about $10^{-14}$ to $10^{-17}$ using the observations of 20 years, and we estimate that about 100 SMBH sources can be detected with the SKA detector.}
   {}


\titlerunning{Determining GW from close galaxy pairs using a BPS }
\authorrunning{J. Liu et al.}

   \maketitle

%

\section{Introduction} \label{1. Introduction}
A gravitational wave (GW), which is described as a space
perturbation of the metric traveling at the speed of light, is a
natural consequence of Einstein's theory of gravity (`general
relativity', Einstein 1916, 1918). It has been accepted (Thorne \&
Braginskii 1976) that supermassive black hole (SMBH) binaries, which
are ubiquitous in the nuclei of low-redshift galaxies (Magorrian et
al. 1998), are boua fide GW sources, with the international Pulsar
Timing Array (PTA) and the proposed Square Kilometer Array (SKA)
(Detweiler 1979; Hellings \& Downs 1983; Kaspi et al. 1994; Wyithe
\& Loeb 2003; Jenet et al. 2005; Yardley et al. 2010). The precision
of the rotation periods for millisecond period pulsars, which is
established via pulse arrival time measurements, allows the
detection of stochastic SMBH GW background spectrum at nHz
frequencies (Rajagopal \& Romani 1995; Jenet et al. 2006). These
observations give us a good prospect for detecting GWs from the
ensemble of these SMBH coalescence events throughout the universe.
Some of these studies have been taken up by other groups over the
years. In particular, the sensitivity curve of the SKA to GWs
emitted by SMBH systems has been obtained from 100 pulsars using the
Australia Telescope National Facility pulsar catalog (Yardley et al.
2010).

The frequent growth of galaxies by merger processes and SMBH in
galaxies result in some obvious mechanisms about the formation of
SMBHs. Several studies have argued that an occupation fraction of
SMBH as low as 0.01 at reshift 5 can be obtained from subsequent
mergers (Menou et al. 2001 and references therein). Spherical
galaxies often lead to coalescence timescales that are longer than
the Hubble time, while highly flattened or triaxial galaxies lead to
faster coalescence (Wyithe \& Loeb 2003). Therefore this expected GW
radiation investigation from coalescing SMBHs depends on the merger
rate of massive galaxies, the demographics of SMBHs at low- and
high- redshift, and the dynamics of SMBH binaries. Meanwhile the
SMBHs need to reach a regime where GW radiation can drive the main
evolutionary process (Begelman et al. 1980). If the galaxy pairs can
evolve to this GW radiation regime, they will gradually lose angular
momentum to GW radiation and will eventually coalescence. The
emission of GWs from SMBH sources requires that the binary coalesce
in less than the Hubble time.

Some results of close galaxy pairs have been released from the Sloan
Digital Sky Survey (SDSS) data (Ellison et al. 2008, 2010), such as
the star formation rate, the merger rate of massive galaxies, the
demographics of SMBHs at low redshift ($z<0.1$) and the dynamics of
SMBHs. A binary population synthesis (BPS) approach has been applied
to study the characteristics of clusters and galaxies (see, e.g.,
Han et al. 2007; Zhang et al. 2010). This method has also been
systematically taken up by some groups to investigate the GW
radiation from compact binaries in the Galaxy (Nelemans et al. 2001,
Liu 2009; Liu et al. 2010; Ruiter et al. 2010). However, the GW
radiation from SMBHs, combined with the BPS model and observation
results, are not well known. Here we report how BPS , using SDSS
results, can be used to determine the GW radiation from SMBHs.

 In the next section, the BPS model is described. In Sect. 3, we present the results of our simulations.
And the discussion is given in Sect. 4.

\section{The binary population synthesis model} \label{2. The binary population synthesis model}

\subsection{Confirming the evolution trace of SMBH systems with OJ287}\label{2. Confirming the evolution trace of SMBH systems with OJ287}

Supermassive black holes are expected to be a type of GW sources,
but few studies can accurately determine a galaxy pair's internal
compositions and kinematics equations. The BL Lacertae object OJ287
(Sillanpaa et al. 1988, 1996; Valtonen 2007) provides a possible
insight into the precessional elliptical orbital evolution through
GW radiation in the phase of coalescence of SMBH binaries. The basic
elements of this phase are as follows: 12--year intervals in optical
outburst, component masses $1.3\times10^{8}\rm M_{\odot}$ and
$1.8\times10^{10}\rm M_{\odot}$, an orbital period of nine years, an
eccentricity of 0.67, a distance of 1.3 {\rm Gpc}.

 In accordance with
Einstein's general relativity, the metric can be written as

\begin{equation}
g_{\alpha \beta}=\eta_{\alpha \beta}+h_{\alpha\beta},
\end{equation}
where $\eta_{\alpha \beta}$ is Minkowsky metric, and
$h_{\alpha\beta}$ stands for GWs with $\mid h_{\alpha\beta}\mid\ll$
1. The amplitude and the luminosity of GW radiation are

\begin{equation}
h^{jk}(t,r)=\frac{2\rm G}{\mathcal {R}\rm c^4}\frac{{\rm
d}^{2}D^{jk}}{{\rm d}t^2}
\end{equation}
and
\begin{equation}
L_{\rm GW}=\frac{\rm G}{45\rm c^5}<\frac{{\rm d}^{3}D^{ij}}{{\rm
d}^{3}t}\frac{{\rm d}^{3}D_{ij}}{{\rm d}^{3}t}>,
\end{equation}
  where G and c are the gravitational constant and speed of light, respectively. $\mathcal {R}$ is the distance from the
  GW sources to the Earth, $D_{ij}$ is the polar moment of mass, and the symbol $<  >$ denotes the averaging.

 The SMBHs more increasingly closer because of GW radiation over time, therefore the lifetime of galaxy pairs
before the final in-spiral is (Jaffe \& Backer 2003)
\begin{equation}
t_{\rm{GW}}= 1.1\times10^{6}\,\rm yr\,{\rm
M}_{8}^{-5/3}\emph{P}_{\rm
orb}^{8/3}\frac{(1+\emph{q})^2}{\emph{q}}(1-\mathit{e^{2}})^{7/2},
\end{equation}
 where $M_{8}$ stands for the total mass of the binary in units of $10^8M_{\odot}$, \emph{q} $(=m_{1}/m_{2}<1 )$ is the mass ratio, $P_{\rm orb}$ is the
  orbital period in years. Meanwhile, the orbital angular momentum loss from two point masses is given by
  \begin{equation}
\frac{{\rm d}J_{\rm{GW}}/{\rm d}t}{J_{\rm{orb}}}=
-8.315\times10^{-10}\frac{m_1m_2(m_{1}+m_{2})}{r^{4}}\frac{1+\frac{7}{8}e^{2}}{(1-e^{2})^{5/2}}\rm
\,yr^{-1},
\end{equation}
 where \emph{r} is the separation of SMBHs in the galaxy pairs.

As displayed above, an important quantity to be considered is the GW
amplitude \emph{h}, here we give a simple description of the
calculation of \emph{ h} in our simulations. First, we give the
definition for the polar moment of mass $D_{ij}$ in Eq. (3)

 \begin{equation}
D_{ij}=q_{ij}-\frac{1}{3}\delta_{ij}\delta^{kl}q_{kl},
\end{equation}
 where $q_{ij}=3\int y^{i}y^{j} \mathbb{T}^{00}(t,y)$, and
 $i,j=1,2$. In general, we need to calculate this energy momentum
 tensor for an isolated system. For our particle binaries we just simulate
 the equation of motion of each BH using a numerical method. The primary of an SMBH system is defined here as the initially
 more massive component ($m_{2}$) throughout. We also
 assumed that the primary BH $m_{2}$ is at the origin of coordinate, and that the
 secondary BH $m_{1}(<m_{2}$) traces out the precessional elliptical
 orbit around this primary.

 According to Eq. (16) of Antonacopoulos (1979), the equation of motion
 of the secondary $m_{1}$ can be described as

 \begin{figure}[h,t]
\centerline{\includegraphics[angle=0,width=0.55\textwidth]{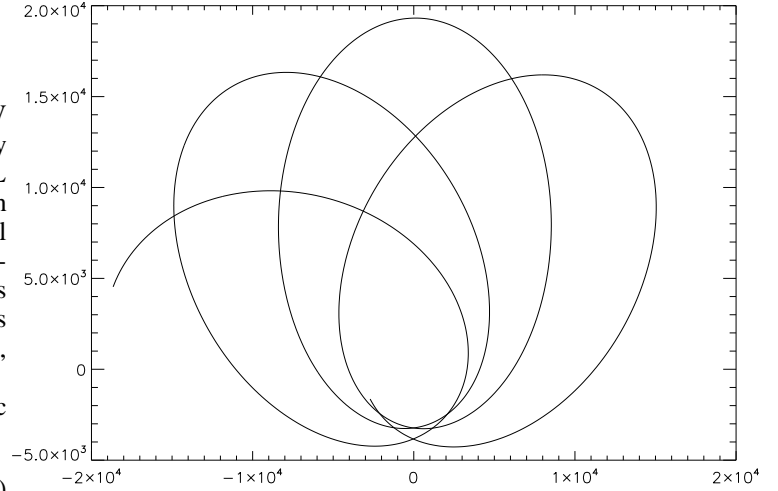}}
\caption{Orbital trace template of SMBH systems using OJ287. The
primary BH is at coordinates (0,0). The precessional elliptical
curve represents the evolutionary trace of the secondary BH.}
\label{fg:dnu}
\end{figure}
 \begin{eqnarray}
\frac{d \bar{u}_{1}}{dt}=-{\rm
G}\frac{m_{2}(\vec{x}_{1}-\vec{x}_{2})}{\mid
\vec{x}_{1}-\vec{x}_{2}\mid ^{3}}\hspace{129pt}\nonumber
\\+\frac{1}{{\rm c^2}}[-{\rm
G}\frac{m_{2}(\vec{x}_{1}-\vec{x}_{2})}{\mid
\vec{x}_{1}-\vec{x}_{2}\mid ^{3}}\vec{u_{1}}^{2}+{4\rm
G}\frac{m_{2}(\vec{x}_{1}-\vec{x}_{2})\cdot \vec{u}_1}{\mid
\vec{x}_{1}-\vec{x}_{2}\mid ^{3}}\vec{u}_1\nonumber\\+{4\rm
G}^{2}\frac{m_{2}^{2}(\vec{x}_{1}-\vec{x}_{2})}{\mid
\vec{x}_{1}-\vec{x}_{2}\mid ^{4}}]+\frac{1}{{\rm c^4}}[-2{\rm
G}^{2}\frac{m_{2}^2(\vec{x}_{1}-\vec{x}_{2})\cdot \vec{u}_1}{\mid
\vec{x}_{1}-\vec{x}_{2}\mid ^{4}}\vec{u_{1}}\nonumber\\-{9\rm
G}^3\frac{m_{2}^3(\vec{x}_{1}-\vec{x}_{2})}{\mid
\vec{x}_{1}-\vec{x}_{2}\mid ^{5}}+{2\rm
G}^{2}\frac{m_{2}^{2}[(\vec{x}_{1}-\vec{x}_{2})\cdot\vec{u}_{1}]^2}{\mid
\vec{x}_{1}-\vec{x}_{2}\mid ^{6}}(\vec{x}_{1}-\vec{x}_{2})],
\end{eqnarray}
where $\vec{x}_{1}$ and $\vec{x}_{2}$ are the vector position of
secondary and primary in the BH binaries, respectively. And
$\vec{u}_{1}$ is the vector velocity of secondary BH. To facilitate
the calculation, we derive from Eq. (7) using the polar coordinate
 \begin{equation}
\ddot{r}-r\dot{\varphi}^{2}=-\frac{{\rm G} m_2}{r^2}+\frac{3{\rm
G}m_2}{{\rm c^2}r^2}-\frac{{\rm G}m_{2}\dot{\varphi}^2}{\rm
c^2}+\frac{4{\rm G^2}m_{2}^2}{{\rm c^2}r^3}-\frac{9{\rm
G^3}m_{2}^3}{{\rm c^4}r^4}
\end{equation}
 \begin{equation}
r\ddot{\varphi}+2\dot{r}\dot{\varphi}=(\frac{4{\rm G}m_2}{\rm
c^2}-\frac{2{\rm G^2}m_{2}^2}{{\rm c^4}r^2})\dot{r}\dot{\varphi},
\end{equation}
here $r=|\vec{x}_{1}- \vec{x}_{2}|$. Here we neglect the influence
of the smaller quantity (such as $\mathcal {O}(\bar{c}^5)$) in Eqs.
(7)-(9). We did not perform the calculation in full relativity.

To fix the orbital trace of SMBHs, which is consistent with OJ 278's
modulation, we investigated the numerical result using the followed
initial parameters of the position of periastron: $r_{0}=3162\rm
AU$, $\varphi_{0}=0$, $\dot{r}_{0}=0$ and the tangent components of
the velocity $v_{0}=0.275\rm c$, the subscript $0$ denotes the
initial values.
 Therefore, Fig. 1 shows an orbital evolution trace of the secondary
 BH. Note that our calculation of the precessing-binary model is slightly different from that of Fig. 1 of Valtonen (2007), which is consistent with
 the fixed optical outbursts of OJ287 according to an accretion disk model.
 In other words, we excluded the influence of the accretion disk from our
 definition. That is mainly because the accretion disk dominates the
 electromagnetic radiation. Our simulations focus on the
 importance of GW radiation.

To analyze the GW amplitude over time, from Eq. (6), the components
of the mass tensor (Capozziello et al. 2008) are

\begin{eqnarray}
D_{11}=m_{1}r^{2}(3\cos^{2}\varphi-1)
\end{eqnarray}

\begin{eqnarray}
D_{12}=3m_{1}r^{2}\cos\varphi \sin\varphi
\end{eqnarray}

\begin{eqnarray}
D_{22}=m_{1}r^{2}(3\sin^{2}\varphi-1)
\end{eqnarray}

\begin{eqnarray}
D_{33}=-m_{1}r^{2}.
\end{eqnarray}
Here other quantities $D_{ij}=0$. The change of the GW amplitude is
illustrated in Fig. 2. We can note four kinds of information from
it. Firstly, we consider OJ287 as our standard template and obtain
some typical feedback: the orbital period is 9.4 yr, the
precessional motion is $40^{\rm o}$ per period, and the eccentricity
is 0.71. Secondly, this numerical curve covers about 40 years from
1996 to 2036. And the maximum quantities of GW amplitude will appear
on the position at which the secondary BH moves to the position of
periastron. To some extent, we also find that the events that
happened in maximum position are consistent with those of the
optical outburst. For example, during 2005--2006 (Valtonen et al.
2008), the time of outburst from OJ287 is very close to a position
of maximum GW radiation in our manifestation. In other words, if the
OJ287 experiences an outburst in 2020, the very high possibility of
GW detection can be investigated using pulsars around 2020. Thirdly,
we find that the GW signal from this type of individual SMBH sources
at the Earth is unlikely to be detected (the position of the symbol
``today" is about $5\times 10^{-16}$ in 2012.) using a current PTA
detector, which agrees with estimates from previous studies (Hobbs
et al. 2009, Sesana, Vecchio \& Volonteri 2009; Yardley et al. 2010
). Finally, these descriptions of the GW are the crucial criteria
that we adopt throughout.

 \begin{figure}[h,t]
\centerline{\includegraphics[angle=0,width=0.55\textwidth]{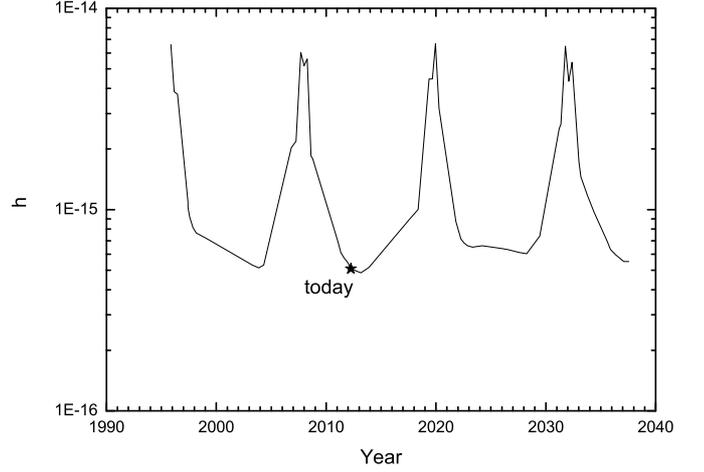}}
\caption{Amplitude \emph{h} as a function of the observation time
for the OJ287's model. The symbol ``$\star$" indicates the position
at which the GW radiation from OJ287 is, $\sim 5\times 10^{-16}$ in
2012.} \label{fg:dnu}
\end{figure}

\subsection{Parameters in the Monte Carlo simulation}\label{2. Parameters in the Monte Carlo simulation}

To systematically investigate the GW radiation of galaxy pair
systems, we performed a Monte Carlo simulation through which we
followed the evolution of a sample of 1 million binaries. The
description of our input physics parameters is given as follows.\\
(i) The properties of the overall galaxy pair samples have been
displayed in the released SDSS data (Ellison et al. 2010), including
projected and angular separation, relative velocities ($\Delta
\upsilon$), stellar mass ratio, and redshift. We present polynomial
curve fitting formulae, which we used to construct the distribution
function of galaxy pairs (see Figs. 1 and 2 of Ellison et al. 2010),
which approximate the evolution of SMBHs for a wide range of mass
\emph{M}($10^{8}-10^{12}M_{\odot}$) and stellar environment $\rm log
\Sigma$($-1.4 - 1.4 \rm Mpc^{-2}$). In our Fig. 3 (see panels
(a)-(g)) we give the fitting curves of the physical properties for
the galaxy pair sample using these released SDSS data.\\
(ii) The star formation rate(SFR) is taken to be constant in the
simulation (Ellison et al. 2008).\\(iii) According to the derivation
of the black hole mass function (BHMF) at low redshift (Tamura et
al. 2006), a simple polynomial curve fitting approximation of BHMF
is also used in panel (h) of Fig. 3.\\(iv) We assumed that the
merger rate of nearby ($z<0.1$) galaxy pairs is 0.2 $\rm yr^{-1}$ in
the phase of coalescence with a chirp mass of $10^{10} \rm M_\odot$.

\begin{figure}[h,t]
\centerline{\includegraphics[angle=0,width=0.52\textwidth]{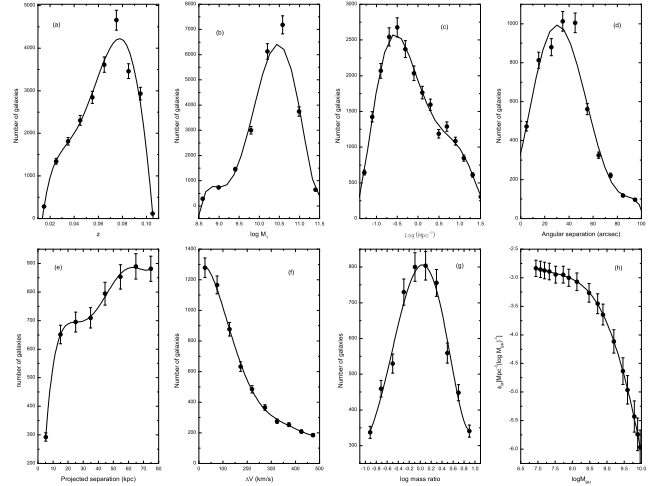}}
\caption{Fitting physical property curves of Monte Carlo parameters
are derived from SDSS (Ellison et al. 2010) in panels (a)--(g),
whose error bars represent the uncertainties of estimation due to
the fitting errors in galaxy pairs of SDSS. The BH mass function
curve is also display in panel (h), which is the same as that of
Tamura et al. (2006).} \label{fg:dnu}
\end{figure}

To construct the SMBH systems, other parameters used in the typical
SMBH
binary formation in this work are given as follows.\\
(i) The correlation between the mass $M_{\rm BH}$ of the BH and the
velocity dispersion $\sigma$ is given by (Tremaine et al. 2002)

\begin{equation}
\rm {log} (\emph{M}_{BH}/M_{\odot})=\alpha+\beta \rm log
(\sigma/\sigma_{0}),
\end{equation}
where $\sigma_{0}=200\rm {km/s}$, $\alpha=8.13$ and $\beta =4.02$.\\
(ii) For the BH mass and the luminosity relation, we adopted the
\emph{B}-band relation described by Bell et al. (2004)
\begin{equation}
\rm {log} (\emph{M}_{BH})=1.19 (\rm log\emph{L}-10.0)+8.18.
\end{equation}
(iii) A gravitational potential in a galaxy includes the bulge,
disk, and halo potentials. The disk and bulge potentials are derived
from Miyamoto \& Nagai (1975)
\begin{equation}
\Phi (R, z)=\frac{GM_{i}}{\sqrt{R^{2}+(a_{i}+\sqrt{z^{2}+b_{i}})}},
\end{equation}
where R and z are the galaxy center coordinate parameters and
$R=\sqrt{x^{2}+y^{2}}$. The index i corresponds to bulge and disk.
The halo potential is given as (Paczynski 1990)
\begin{equation}
\Phi (r)=-\frac{GM_{c}}{r_{c}}[\frac{1}{2}{\rm ln}(1+\frac{r^2}{
r_{c}^2})+\frac{ r_c}{r}{\rm atan}(\frac{r}{r_c})].
\end{equation}
A more detailed description of the parameters in Eqs. (16) and (17)
can be found in Sect. 2 of Paczynski (1990). We obtained the
eccentricity of the SMBH samples from the random number ($e=x, 0\leq
x \leq1$). Throughout this paper, we adopted the cosmological model
with $H_{0}=70\rm kms^{-1}Mpc^{-1}$, $\Omega _{\rm M}=0.3$ and
$\Omega_{\rm \Lambda}=0.7$.

\section{Results}\label{3. Results}
With a BPS simulation we obtained a series of SMBHB systems from a
sample of 1 million initial SMBHBs. Based on the sensitivity curves
of the PTA and SKA detectors (Yardley et al. 2010), we created the
time series for the source signals and added the individual time
series of the calculated sources to produce the total data stream.
Similar calculations of the GW signal analysis combined with a GW
detector can be found in Sesana, Vecchio \& Colacino (2008), Liu
(2009), and Ruiter et al. (2010).

Figs. 4 and 5 show the variation of the strain amplitude and
luminosity with time from the GW radiation of SMBHs during a long
time observation (e.g. 20 years) in our Monte Carlo simulations.
These tendency changes over time reveal that the GW signals obtained
by current and planned detectors should be variable events. They
will likely detect two higher GW radiation signals around 2020 and
2032, which is consistent with the predicted time of optical bursts
of a merging galaxy (such as OJ287, Valtonen 2007). This is because
we adopted a precession of orbital evolution of OJ287 as our orbital
evolutionary model to investigate the orbital track changes caused
by GW radiation in the simulations. In other words, according to our
descriptions in Sect. 2, the maximum outbursts of GW radiation and
the optical outburst are happening at the exact time when the
secondary BH crashes the periastron of the primary BH.

Fig. 4 shows that the strain amplitude of GW radiation in the range
of $4\times10^{-17}$ to $7\times10^{-14}$  will be detected by the
SKA detector (Yardley et al. 2010). From Fig. 5 we can see that our
calculated MBHB samples can radiate a high luminosity due to GWs of
between $ \sim 1\times10^{48}$ and $10^{51}$ erg$\cdot $s$^{-1}$,
which possibly means that the influence of GW dominates the total
energy loss of the galaxy pairs in the early phase of coalescence.
Meanwhile, this variation of the strain amplitude or luminosity with
time maybe provides an indirect evidence for a GW. Especially using
20 years of observational time, the pulsar timing measurements will
confirm a period variation in residual data.

\begin{figure}[h,t]
\centerline{\includegraphics[angle=0,width=0.55\textwidth]{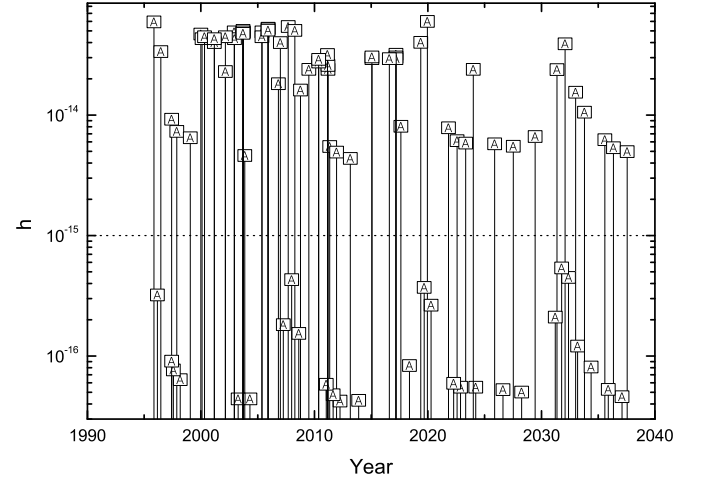}}
\caption{Strain amplitude of GW radiation variation with time. The
symbol ``A" represents the average value position of the summed GW
strain in each bin. The dotted line (log $ h =-15$) marks the
highest sensitivity of SKA.} \label{fg:dnu}
\end{figure}

\begin{figure}[h,t]
\centerline{\includegraphics[angle=0,width=0.58\textwidth]{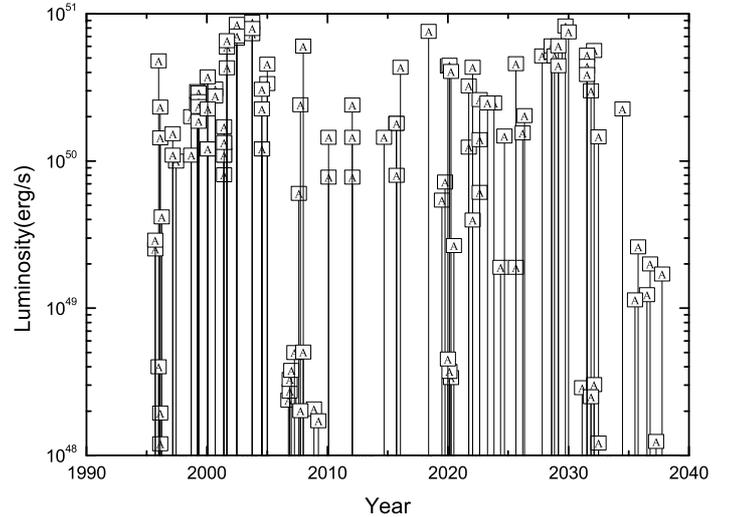}}
\caption{Similar to Fig. 4, but for the luminosity variation of GW
radiations with time.} \label{fg:decayres}
\end{figure}

\begin{figure}[h,t]
\centerline{\includegraphics[angle=0,width=0.58\textwidth]{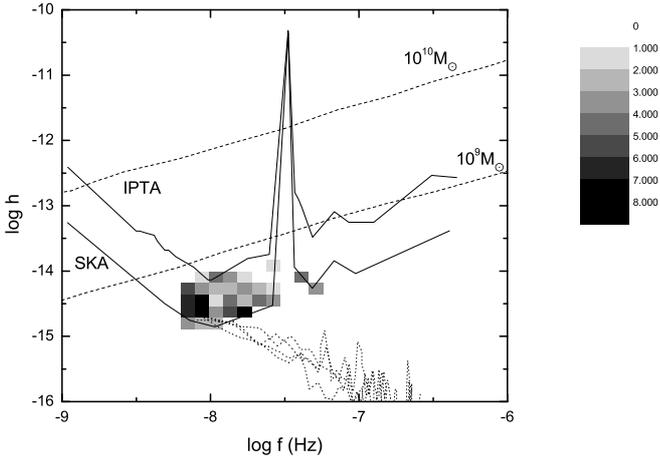}}
\caption{Strain amplitude \emph{h} as a function of the frequency
for the BPS simulations of individual resolvable MBHBs. The gray
scale gives the density distribution of the resolvable systems.
Other lines are cited by the sensitivity studies (Sesana, Vecchio \&
Colacino 2008; Yardley et al. 2010) using the PTA detector. }
\label{fg:allres}
\end{figure}

\begin{table}
 \begin{minipage}{90mm}
 \caption{The percentage of detectable sources with different eccentricities (e).
 Note that the total number of detectable SMBH binaries in our model is 100.}
  \tabcolsep=0.03in
 \label{tab1}
   \begin{tabular}{|l|c|c|c|c|}
\hline
 eccentricities &$0<e<0.2$ &$0.2<e<0.5$&$0.5<e<0.7$&$0.7<e<1.0$\\
\hline
percentages &7.0\% &18.0\%&33.0\%&42.0\% \\
\hline
\end{tabular}

\end{minipage}

\end{table}

Using a 20 year-integration time, we plot in Fig.6 the GW radiation
distribution from the resolved MBHB systems in our simulation. This
shows that the predicted GW radiation is below the PTA sensitivity
curve. Therefore the individual SMBH sources cannot radiate
sufficient GW signals for the current detector. However, these
signals will be detected by SKA, the detectable number of SMBHs is
about 100, which is above the SKA detection threshold. Our
prediction agrees well with that of Sesana, Vecchio \& Colacino
(2008), Sesana, Vecchio \& Volonteri (2009), and Yardley et al.
(2010) (see the discussion).

Iwasawae al. (2010) have systematically investigated the importance
of eccentricity in SMBH binaries due to GW radiation. For the 100
detectable sources in Fig.6, we also give the percentages of these
sources for different eccentricities using Table 1. From it we find
that the low-eccentricity binaries cannot dominate the detectable
sources. The table also implies that once the individual SMBH
binaries become highly eccentric, the GW radiation should be
observable by the SKA detector. Meanwhile, we find for these
detectable sources that the average of chirp mass $\langle
\mathcal{M}_{\rm{chirp}}\rangle
=\langle(m_{1}m_{2})^{3/5}(m_{1}+m_{2})^{-1/5} \rangle$ is about
$8.9\times10^{10}M_{\odot}$.

\section{Discussion}\label{4. Discussion}
We have shown that there are good reasons to expect GW radiation
from SMBHs and that some amplitudes will be measured by SKA in the
range of $10^{-14}<h<10^{-15}$. The results from a BPS approach
using Monte Carlo simulations are innovative in GW radiation
studies, which also regulate the calculation together with the
analytical fitting functions of observation results from SDSS. In
these close galaxy samples, very close SMBH binaries may produce
very strong GW signals in orbital motion processes, which may be
reflected by the profiles of GW forms. Especially the SMMB binaries
with short orbital periods ($<$10 yr) may even be detectable by
photometric and high-energy radiation analysis using ground-based or
space telescopes. Note that the precession of orbital period caused
by GW radiation from highly eccentric SMBH binaries can influence
the evolution of galaxy pairs.

To study the SMBH systems such as GW sources, we needed some basic
parameters for the SMBH systems (the primary BH mass $m_{2}$, the
secondary BH mass $m_{1}$, the orbital period $P_{\rm orb}$, and so
on), which were derived from the distributions of SDSS data (Ellison
et al. 2008, 2010) using the polynomial curve fitting approximation.
First, few studies focus on the GW radiation from SMBH systems
together with the latest findings of observations, although our
estimation methods are straightforward. Second, indeed, enough close
galaxy pairs have been released from the SDSS data, but the error
(or gray) bars (see Fig. 1 of Tamura et al. 2006 and Fig. 5 of
Ellison et al. 2008) caused by the observations still exist in our
simulation. Therefore a maximum ratio with respect to observations
is approximately 10, which is derived from the maximum difference
obtained from error bars. Third, for the redshift evolution of SMBH
systems, we investigated the redshift distribution up to $z\sim0.1$,
which compares very well with a sky-averaged constraint on the
merger rate of nearby SMBH sytems using a pulsar timing array
measurement, implied by the low-redshift distance approximation
$r=cz/H_{0}$.

The main purpose of this study is to find out how many resolved SMBH
sources we can explore with a pulsar timing detector from the GW
radiation. Given the shape of the sky- and polarization-averaged
sensitivity curve of the PTA (Yardley et al. 2010), the sensitivity
curve is useful for comparing the strain amplitude of the GWs from
an SMBH system with the average noise to see whether the individual
sources can be detected by the PTA or SKA detector. A resolved
source with frequency $f$ and strain amplitude $h$ that is observed
over a time $T_{\rm obs}$ (=20 years) will appear in the Fourier
spectrum of the data as a single spectral line. Therefore the SMBH
systems, which are the only ones in their corresponding periodogram
ranges ($\Delta f=1/T_{\rm obs}=1.6\times10^{-9}$) and have higher
strains than the sensitivity noise curve of the detector, are called
resolved sources.

As we displayed in Fig. 6, the significantly different results can
be explained as follows. Firstly, there are individual resolved SMBH
systems in our nearby galaxy pairs above the predicted stochastic GW
background, which is likely to be detected using the PTA in the near
future (Sesana, Vecchio \& Colacino al. 2008). Note that our
estimation of GW radiation agrees with that of Sesana, Vecchio \&
Colacino (2008), although these authors used different methods for
the underlying SMBH population, for instance, we used a precession
of orbital period evolution of OJ287 and a series of derived
observation results from SDSS in our simulation. Secondly, the
distance was obtained using $D=cz/H_0$, which is an acceptable
approximation for relatively low redshift of galaxy pairs (Davis \&
Lineweaver 2004). All resolved sources in Fig.6 are located at
fairly low redshift ($z<0.1$), and not at $0.2<z<1.5$, which is one
condition of Sesana, Vecchio \& Volonteri (2009), who predicted that
their individual SMBH is below the current sensitivity. Accordingly,
the major difference in the estimating the resolved sources between
our model and the model of Sesana, Vecchio \& Volonteri (2009)
arises from the distance \emph{D}. Thirdly, Iwasawa et al. (2011)
has pointed out that the eccentricity of most of SMBH binaries can
result in a rapid merger through GW radiation. The 75\% detectable
sources in Fig. 6 show high eccentricities, but Yarledy et al.
(2010) did not consider the eccentric SMBH binaries. Consequently,
the influence of the distance and eccentricity are the main
difference from other papers.

If the orbital motion of the SMBH system is appropriately reproduced
using PN2 methods, we expect that the maximum GW radiation should be
associated with the precise evolution trace of each BH component
movement in the full relativity field. This is because the secondary
BH crashes the periastron, which leads to the outburst GW signal.
Furthermore, based on our estimate of the merger rate of close
galaxies and making several evolutionary hypotheses, we can compute
the number of SMBH sources that can be detected by the PTA (or SKA)
detector. Additionally, the strength of GW signal from the merger of
SMBH binaries will be blurred by other fierce energy release
processes (such as supernova explosion, Mazzali et al. 2008).
Finally, the evolutionary properties of SMBHs could indeed be more
complex than we have described. For example, the mass transfer
process between the components could be important and may even
account for quasars. Meanwhile, the merging process may occur more
rapidly than the evolution of SMBH, and even more BHs may reside in
the core of galaxy.

To summarize, a numerical simulation of the GW radiation from the
nearby galaxy pairs has been carried out using OJ287's modulation.
We used the result of SDSS data and several notable conclusions from
the literature to follow the SMBH evolution using a BPS approach. In
addition to a GW detector (such as SKA), we used these methods to
examine the GW variation with time and estimated the number of
detectable SMBH sources using observations made over 20 years. This
study demonstrates the significant contribution to the GW radiation
from the individual SMBH sources.
\begin{acknowledgements}
We gratefully acknowledge the BPS (Zhanwen Han and Fenghui Zhang)
group at Yunnan Observatory for a discussion. We especially thank
the referee for useful comments and improving of the manuscript.
This work is supported by the program of the light in China's
Western Region (LCWR) (No. XBBS201022), Natural Science Foundation
(No. 11103054) and Xinjiang Natural Science Foundation (No.
2011211A104). This project/publication was made possible through the
support of a grant from the John Templeton Foundation. The opinions
expressed in this publication are those of the authors and do not
necessarily reflect the view of the John Templeton Foundation. The
funds from John Templeton Foundation were awarded in a grant to The
University of Chicago, which also managed the program together
National Astronomical Observatories, Chinese Academy of Sciences
(No. 100020101).

\end{acknowledgements}

\clearpage


\begin{thebibliography}{}\label{thebibliography}




\bibitem[\protect\citeauthoryear{Antonacopoulos}{1979}] {anto79} Antonacopoulos, G. 1979, Ap\&SS, 62, 217

\bibitem[\protect\citeauthoryear{Begelman et al.}{1979}] {Bege79} Begelman, M. C., Blandford, R. D., \& Rees, M. J. 1980, Nature, 287, 307

\bibitem[\protect\citeauthoryear{Bell et al.}{2004}] {Bell04} Bell, E. F., et al. 2004, ApJ, 608, 752


\bibitem[\protect\citeauthoryear{Capozziello et al.}{2008}] {Capo04} Capozziello S., et al. 2008, Mod. Phys. Lett., 23, 99

\bibitem[\protect\citeauthoryear{Davis}{2004}] {Davi04} Davis, T. M., \& Lineweaver C. H., 2004, Publ. Astron. Soc. Aust., 21, 97

\bibitem[\protect\citeauthoryear{Detweiler}{1979}] {detw79} Detweiler, S. 1979, ApJ, 234, 1100

\bibitem[\protect\citeauthoryear{Einstein}{1916}]{eins16} Einstein, A. 1916, Preuss. Akad. Wiss. Berlin, Sitzungsberichte der
physikalisch--mathematischen Klasse (Berlin: Springer), 688

\bibitem[\protect\citeauthoryear{Einstein}{1918}]{eins16} Einstein, A. 1918, Preuss. Akad. Wiss. Berlin, Sitzungsberichte der
physikalisch--mathematischen Klasse (Berlin: Springer), 154

\bibitem[\protect\citeauthoryear{Ellison et al.}{2008}] {elli08} Ellison, S. L., Patton, D. R., Simard, L., \& McConnachie, A. W. 2008, AJ, 135, 1877

\bibitem[\protect\citeauthoryear{Ellison et al.}{2010}] {elli10} Ellison, S. L., Patton, D. R., Simard, L., McConnachie, A. W., Baldry, I. K.; \& Mendel, J. T. 2010, MNRAS, 407, 1514

\bibitem[\protect\citeauthoryear{Han et al.}{2007}] {Han07} Han, Z., Podsiadlowski, Ph., \& Lynas-Gray, A. E. 2007, MNRAS, 380, 1098

\bibitem[\protect\citeauthoryear{Hellings \& Downs}{1983}] {hell83} Hellings, R. W., \& Downs, G. S. 1983, ApJ, 265, 39

\bibitem[\protect\citeauthoryear{Hobbs et al.}{2009}] {hobb09} Hobbs, G.B., et al. 2009, PASA, 26, 103
\bibitem[\protect\citeauthoryear{Liu}{2009}] {liu09} Liu, J. 2009, MNRAS, 400, 1850
\bibitem[\protect\citeauthoryear{Liu et al.}{2010}] {liu10} Liu, J., Han, Z., Zhang, F., \& Zhang, Y. 2010, ApJ, 719, 1546
\bibitem[\protect\citeauthoryear{Iwasawa et al.}{2011}] {Iwas11} Iwasawa, M., et al. 2011, ApJL, 731, 9
\bibitem[\protect\citeauthoryear{Jaffe \& Backer}{2003}] {Jaff03} Jaffe, A. H., Backer, D. C. 2003, ApJ, 583, 616

\bibitem[\protect\citeauthoryear{Jenet et al.}{2005}] {jen05} Jenet, F. A., Creighton, T., \& Lommen, A. 2005, ApJ, 627, 125


\bibitem[\protect\citeauthoryear{Jenet et al.}{2006}] {jen06} Jenet, F. A., et al. 2006, ApJ, 653, 1571

\bibitem[\protect\citeauthoryear{Kaspi et al.}{1994}] {kasp94} Kaspi, V. M., Taylor, J. H., \& Ryba, M. F. 1994, ApJ, 428, 713

\bibitem[\protect\citeauthoryear{Magorrian et al.}{1998}] {mag98} Magorrian, J., et al. 1998, AJ, 115, 2285


\bibitem[\protect\citeauthoryear{Mazzali et al.}{2008}] {maz98} Mazzali, P. A.., et al. 2008, Science, 321, 1185

\bibitem[\protect\citeauthoryear{Menou et al.}{2001}] {meno01} Menou, K., Haiman, Z., \& Narayanan, V. K. 2001, ApJ, 558, 535

\bibitem[\protect\citeauthoryear{Miya et al.}{1975}] {miya75} Miyamoto, M., \& Nagai, R. 1975, PASJ, 27, 533

\bibitem[\protect\citeauthoryear{Nelemans et al.}{2001}]{nele01} Nelemans, G., Yungelson, L. R., \& Portegies-Zwart, S. F., 2001, A\&A, 375, 890


\bibitem[\protect\citeauthoryear{Paczynski}{1990}] {Pacz90} Paczynski, B., 1990, ApJ, 348, 485
\bibitem[\protect\citeauthoryear{Rajagopal \& Romani}{1995}] {Raja95} Rajagopal, M., \& Romani, R. 1995, ApJ, 446, 543

\bibitem[\protect\citeauthoryear{Ruiter et al.}{2010}] {Ruit10} Ruiter, A. J., Belczynski, K., Benacquista, M., Larson, S. L., \& Williams, G., B., 2010, ApJ, 717, 1006
\bibitem[\protect\citeauthoryear{Sesana et al.}{2008}] {sesa08} Sesana, A., Vecchio, A., \& Colacino, C. N. 2008, MNRAS, 309, 192

\bibitem[\protect\citeauthoryear{Sesana et al.}{2009}] {sesa09} Sesana, A., Vecchio, A., \& Volonteri, M. 2009, MNRAS, 394, 2255

\bibitem[\protect\citeauthoryear{Sillanpaa et al.}{1988}] {sill88} Sillanpaa, A., Haarala, S., Valtonen, M. J., Sundelius, B., \& Byrd, G. G., 1988, ApJ, 325, 628
\bibitem[\protect\citeauthoryear{Sillanpaa et al.}{1996}] {sill96} Sillanpaa et al., 1996, A\&A, 305, 17

\bibitem[\protect\citeauthoryear{Tamura et al.}{2006}] {tam06}Tamura, N., Ohta, K., \& Ueda, Y. 2006, MNRAS, 365, 134
\bibitem[\protect\citeauthoryear{Thorne \& Braginskii}{1976}] {thor76} Thorne, K. S., \& Braginskii, V. B. 1976, ApJ, 204, L1

\bibitem[\protect\citeauthoryear{Tremaine et al.}{2002}]{trem02} Tremaine, S., Gebhardt, K., Bender, R., et al. 2002, ApJ, 574, 740
\bibitem[\protect\citeauthoryear{Valtonen}{2007}] {valt07} Valtonen, M. J. 2007, ApJ, 659, 1074
\bibitem[\protect\citeauthoryear{Valtonen}{2008}] {valt08} Valtonen, M. J., Kidger, M., Lehto, H., \& Poyner G. 2008, A\&A, 477, 407
\bibitem[\protect\citeauthoryear{Wyithe \& Loeb}{2003}] {wyi03} Wyithe, J. S. B., \& Loeb, A. 2003, ApJ, 590, 691

\bibitem[\protect\citeauthoryear{Yardley et al.}{2010}] {yard10} Yardley, D. R. B., et al. 2010, MNRAS, 407, 669





\bibitem[\protect\citeauthoryear{Zhang et al.}{2010}] {Zha11} Zhang, F., Han, Z., Li, L., Shan, H., \& Zhang, Y., 2010, MNRAS, 408, 1283
\end{thebibliography}
\end{document}